\documentclass[twocolumn,amsmath,amssymb,groupedaddress, nofootinbib]{revtex4-1}

\usepackage{graphicx}
\usepackage{amssymb}
\usepackage{bbold}
\usepackage{xcolor}

\usepackage{hyperref}

\begin{document}
\title{Thermally activated flow in models of amorphous solids}
\author{Marko Popovi{\' c}}
\author{Tom W. J. de Geus}
\author{Wencheng Ji}
\author{Matthieu Wyart}
\address{Institute of physics, EPFL, Lausanne}
\begin{abstract}
Amorphous solids yield at a critical value $\Sigma_c$ of the imposed stress $\Sigma$ through a dynamical phase transition. While sharp in athermal systems, the presence of thermal fluctuations leads to the rounding of the transition and thermally activated flow even below $\Sigma_c$. Here, we study the steady state thermal flow of amorphous solids using a mesoscopic elasto-plastic model. In the H{\' e}braud-Lequex (HL) model we provide an analytical solution of the thermally activated flow at low temperature. We then propose a general scaling law that also describes the transition rounding. Finally, we find that the scaling law holds in numerical simulations of the HL model, a 2D elasto-plastic model, and in previously published molecular dynamics simulations of 2D Lennard-Jones glass.
  \end{abstract}
\maketitle

 \section{Introduction}
 Amorphous solids as diverse as metallic glasses, colloidal glasses, emulsions, foams, and granular matter, exhibit a finite yield stress $\Sigma_c$ beyond which they begin to flow. In athermal systems, this corresponds to a sharp yielding transition, separating solid and fluid phases, which has been extensively studied \cite{Bonn2017, Nicolas2018}. At finite temperature, the transition is rounded by thermally activated flow and becomes a smooth crossover. Understanding the properties of the thermally activated flow is a problem of both fundamental and practical importance.

 Plastic deformation of amorphous solids proceeds through localized plastic events \cite{Argon1979, Falk1998, Schall2007}. Each plastic event produces a localized non-affine strain field which redistributes stresses in the material \cite{Maloney2006}. The ensuing dynamics has been described on a mesoscopic scale by shear transformation zone theory \cite{Falk1998} and soft glassy rheology \cite{Sollich1997}. In these approaches, the mechanical noise produced by stress redistribution of individual events is described by an effective temperature. A different mesoscopic approach, the elasto-plastic model, accounts for the stress redistribution induced by a local rearrangement \cite{Lin2014a, Nicolas2018}. In this model, yielding is a dynamical phase transition. The central quantity describing the system is the density $P(x)$ of regions about to undergo a plastic event, with $x$ denoting the additional stress required locally to trigger an event. In the solid phase $P(x)$ is singular implying system spanning avalanches of plastic events \cite{Mueller2015, Lin2015}, consistent with the observations in numerically simulated amorphous solids \cite{Lemaitre2009, Maloney09, Karmakar2010}.  The flowing phase exhibits non-linear rheology with a diverging correlation length as $\Sigma_c$ is approached from above \cite{Lin2014}.

 This phenomenology is similar to the one found in the depinning transition, where an elastic sheet is driven by a force density $F$ through a disordered potential, and a critical value $F_c$ separates moving and static states \cite{Fisher1998, Giamarchi2006}. However, unlike for the yielding transition, stress redistribution after a depinning event is destabilizing everywhere, which leads to a non-singular $P(x)$ and different exponents characterising the critical behavior close to the transition \cite{Lin2015}.
At finite temperature \cite{Fisher1985, Middleton1992, Chen1995, Nowak1998, Roters1999, Vandembroucq2004, Bustingorry2008, Bustingorry2012}  a scaling law for the interface velocity $V \sim T^{\psi}g((F - F_c)/T^{\psi/\beta})$ was proposed by Fisher \cite{Fisher1985} in the context of charge density wave, where $\psi$ is the transition rounding exponent and $\beta$ is the athermal flow exponent $V \sim (F - F_c)^{\beta}$. It was further argued that $\psi= \beta/\alpha$ \cite{Middleton1992}, where $\alpha$ is a parameter characterising the disordered potential ($\alpha= 1.5$ for smooth potentials), which is supported by simulations \cite{Middleton1992, Chen1995}. On the other hand, numerical investigations of elastic string depinning \cite{Bustingorry2008, Bustingorry2012} found 
a different value of the rounding exponent. Furthermore, in \cite{Purrello2017} the measured steady state flow was found not to follow the scaling law from \cite{Fisher1985}\footnote{Instead, an alternative scaling law was proposed from which follows a logarithmic correction to the rounding exponent. However, the latter is derived by considering the limit $T \to 0$ in a finite system and thus may not hold in the thermodynamic limit for small $T$.}, which was further supported by analysis of elastic line depinning in a washboard potential \cite{Kolton2020}.

The thermal rounding of the yielding transition has been much less studied. In ref. \cite{Chattoraj2010} it was proposed that the thermal fluctuations can be incorporated in the athermal steady state flow $\dot{\gamma} \sim (\Sigma - \Sigma_c)^{\beta}$ as an additive, strain rate dependent, correction of local yield stresses. This approach, supported by molecular dynamics simulations, predicts an exponentially suppressed steady state strain rate $\ln \dot{\gamma} \sim (\Sigma_c - \Sigma)^{3/2}/T$ for $\Sigma_c - \Sigma \gg T^{2/3}$. Here, the exponent $3/2$ is a particular value of $\alpha$ for a smooth disordered potential. In this regime, a gap is found in distribution $P(x)$, and the assumption of additivity corresponds to assuming that the gap size is proportional to $\Sigma_c - \Sigma$.  
In the regime $\Sigma - \Sigma_c \gg T^{2/3}$ flow is dominated by the athermal component and thus well described by construction. It is interesting to note that the rounding exponent at $\Sigma = \Sigma_c$ is consistent with the prediction of ref. \cite{Middleton1992}, up to a logarithmic correction. However, in this approach, the influence of elastic interactions on the distribution $P(x)$ is not considered.

In this work, we study the thermal flow of amorphous solids for different values of the parameter $\alpha$. In particular, we first study the thermal steady state flow and $P(x)$ in a H{\' e}braud-Lequeux (HL) model \cite{Hebraud98, Agoritsas2017} which is a mean-field version of elasto-plastic model with a Gaussian mechanical noise. (Note that here we do not consider the mean-field elasto-plastic model \cite{Lin2016a} which preserves the fat-tails in mechanical noise distribution found in finite-dimensional elasto-plastic model, where our scaling analysis should hold but for which we do not have analytical solutions). We derive analytical expressions for both steady state flow and $P(x)$ in the limit $T\to 0$ and verify these results numerically. We find that the strain rate in the HL model can be written in the scaling form as proposed by Fisher \cite{Fisher1985} and Middleton \cite{Middleton1992}. Finally, we propose that this scaling form holds in finite dimensions with a particular form of the scaling function in the regime $\Sigma_c - \Sigma \gg T^{1/\alpha}$ and test it in the HL model, a two-dimensional elasto-plastic model, and 
molecular dynamics simulations available in the literature \cite{Chattoraj2010}. 

\section{Elasto-plastic model at finite temperature}
Elasto-plastic models aim to capture mesoscopic features of yielding in amorphous solids
\cite{Baret02, Picard05,Nicolas2018}.
The system is divided into $N$ mesoscopic blocks that are larger than localized plastic events. A block $i$ is characterized by the local stress component $\sigma_i$ along the external loading direction, the shear elastic modulus $\kappa$, and a local yield stress $\sigma_{Y, i}$. We express all stresses in units of $\kappa$ and choose $\sigma_{Y,i}$ to be narrowly distributed around $1$ (see the Appendix). \footnote{We expect this choice not to affect the universal properties studied here, as they should not change with the choice of microscopic parameters.}.

In athermal systems, the block $i$ fails when 
$|\sigma_i| \geq \sigma_{Y, i}$. Then, over a time $\tau$ the local stress is decreased by an amount $\delta\sigma_i$ which in our numerical implementation is equal to $\sigma_i$ up to a small random term, see the Appendix. The stress in the rest of the system is redistributed according to an elastic force dipole propagator $\delta\sigma_iG(\vec{r})$, where $\vec{r}$ is the distance from the failing block. 

To study thermal plastic flow we introduce a possibility of thermal activation when $|\sigma_i| < \sigma_{Y, i}$. To each block we assign a potential barrier $E_i= c_E x_{i}^{\alpha}$, where $x_i \equiv \sigma_{Y, i} - \sigma_i$ and $c_E > 0$. 
In a system with a smooth disorder potential a plastic event corresponds to saddle-node bifurcation and  $\alpha = 1.5$. We also consider values $\alpha= 2$ and $\alpha= 1$ corresponding to parabolic and linear potentials with a cusp at the instability \cite{Purrello2017}. The failure probability for the block $i$ with $x_i > 0$ is proportional to $\exp{(-c_E x_i^{\alpha}/T)}$, using units where $k_B= 1$. 
The imposed shear stress in the system sets the average block stress $\Sigma\equiv \sum_i \sigma_i/N= 1 - \sum_i x_i/N $. Finally, the plastic strain rate is the sum of rates over individual plastic events  $\dot{\gamma}\equiv \sum_{i}p_i \delta \sigma_i / (\tau N)$, where $p_i=1$ as long as the block is failing, while $p_i= 0$ otherwise. In the yielding regime $|\Sigma_c - \Sigma| \ll \Sigma_c$, at low strain rates blocks fail at $\sigma_i \approx 1$ and thus $\delta\sigma_i \approx 1$ so that the plastic strain rate can be approximated by the rate of plastic events $\dot{\gamma} \approx \sum_i p_i/(\tau N)$. 

\section{Flow  in H{\' e}braud-Lequeux model}
\subsection{Framework}
We introduce an activated version of the H{\' e}braud-Lequeux model where the state of the system is fully described by the density $P(x, t)$ whose dynamics follows:
\begin{align}
  \label{eq:HLthermal}
  \begin{split}
    \partial_t P(x,t; T)&= \dot{\gamma}\left[D \partial_x^2P(x,t; T) + v\partial_xP(x,t; T) + \delta\left(x-1  \right)\right] \\
    &- \left[\frac{1}{\tau}\Theta(-x) + \frac{1}{\tau}e^{- \frac{x^{\alpha}}{T}} \Theta(x)\right]P(x,t; T) \quad .
  \end{split}
\end{align}
Here, $\Theta$ is the Heaviside theta function, the diffusion constant $D$ characterises the Gaussian mechanical noise experienced by the system after each plastic event, the drift velocity $v$ accounts for the externally controlled stress loading, stress relaxation after a failure is described by the delta function term \footnote{This corresponds to $\delta\sigma_i= \sigma_i$ at each block failure, with the choice $\sigma_{Y, i}= 1$.}. The last two terms account for athermal and thermally activated block failure, respectively, where we have set potential barrier constant $c_E= 1$. In a driven system it is very unlikely for a block to fail with $\sigma_i < -\sigma_Y $ and we neglect this contribution \footnote{For a block to reach $x= 2$ starting from $x= 1$ it has to diffuse distance $\Delta x= 1$ against the imposed stress. Common values \cite{Lin2014a} are $D = 0.18$ and $v \approx 1$ so only $\exp{(-v/D)} \approx 0.004$ of blocks that start from $x= 1$ reach $x= 2$.}.

The system stress is given by (using that $\sigma= 1 - x$):
\begin{align}
\label{eq:stress}
  \Sigma &= 1 - \int\limits_{-\infty}^{\infty}xP(x)dx \quad .
\end{align}
In this work we consider only steady state flows and the strain rate is equal to the plastic strain rate:
\begin{align}
\label{eq:1}
\dot{\gamma} &= \frac{1}{\tau}\int\limits_{-\infty}^{\infty} P(x) \left(\Theta(-x) + \Theta(x)e^{-\frac{x^{\alpha}}{T}}\right) dx \quad .
\end{align}

\subsection{Gap at $T= 0$}

The full solution of Eq. \ref{eq:HLthermal} is in general not available. However, we can calculate the strain rate $\dot{\gamma}$ for ${\Sigma_c -\Sigma \ll \Sigma_c }$ and $T^{1/\alpha} \ll \Sigma_c - \Sigma$.
Below $\Sigma_c$ there is no flow in absence of temperature and therefore plastic events mainly occur by thermal activation. Therefore, in the limit $T \to 0$, we expect an Arrehnius type of flow $\dot{\gamma} \sim \exp{(-A/T)}$, with $A >0$. Given this assumption we show that a gap appears in $P_0(x) \equiv \lim_{T\to 0}P(x)$ by considering the steady state of Eq. \ref{eq:HLthermal}:
\begin{align}
  \begin{split}
\label{eq:33}
  0 = &D \partial_x^2P(x) + v\partial_x P(x) + \delta\left(x - 1\right)\\
  &- \frac{1}{\tau \dot{\gamma}}\left[ \Theta(-x) + e^{-\frac{x^{\alpha}}{T}}\Theta(x)\right]P(x)
  \end{split}
\end{align}
In the limit $T \to 0$, for $x< A^{1/\alpha}$ the relative failure rate $\exp{(-x^{\alpha}/T)}/\dot{\gamma}$ diverges and  and for $x > A^{1/\alpha}$ it vanishes. Therefore, the point $x_c= A^{1/\alpha}$ acts as an absorbing boundary. For $x > x_c$, $P_0(x)$ satisfies:
\begin{align}
  \label{eq:13}
\begin{split}
0 &= D \partial_x^2P_0(x) + v_c\partial_xP_0(x) + \delta\left(x-1  \right) \quad .
\end{split}
\end{align}

The solution of Eq. \ref{eq:13} is:
\begin{align}
  \label{eq:14}
  \begin{split}
  P_0(x)&= \frac{1}{v_c}\left( 1 - e^{-\frac{v_c}{D}(x - x_c)} \right)\Theta(x - x_c) \Theta(1 - x) \\
  &+ \frac{1}{v_c}\left( 1 - e^{-\frac{v_c (1 - x_c)}{D}} \right)e^{-\frac{v_c}{D}(x-1)}\Theta(x - 1) \quad .
  \end{split}
\end{align}
Normalisation of $P_0(x)$ requires $x_c= 1 - v_c$ and thus we can express the gap size $x_c$ in terms of the stress $\Sigma$ by evaluating Eq. \ref{eq:stress} in the limit $T \to 0$:
\begin{align}
  \begin{split}
\label{eq:18} 
  \Sigma & = 1 - \int\limits_{-\infty}^{\infty}xP_0(x) dx=  \frac{1}{2}v_c - \frac{D}{v_c} \quad ,
   \end{split}
\end{align}
and we find:
\begin{align}
\label{eq:30}
x_c&= 1 - \Sigma - \sqrt{\Sigma^2 + 2 D} \quad. 
\end{align}
Since $x_c \to 0$ as $\Sigma \to \Sigma_c$ we have $\Sigma_c= 1/2 - D$ so that for $(\Sigma_c - \Sigma) \ll \Sigma_c$:
\begin{align}
\label{eq:25}
x_c&\simeq \frac{ \Sigma_c -\Sigma}{\frac{1}{2} + D} \quad .
\end{align}

\subsection{Thermal rounding of $P(x)$}
At a small but finite temperature the activation occurs in a region around $x_c$ of a width vanishing with $T$.
To find an approximation of $P(x)$ we linearise the potential barrier $E(x)$ around $x_c$ and look for a solution of 
\begin{align}
\label{eq:2}
 D \partial_x^2P(x) + v\partial_x P(x)  - \frac{1}{\tau\dot{\gamma}}e^{-\frac{x_c^{\alpha}}{T}}e^{-\frac{\alpha x_c^{\alpha-1}}{T} (x - x_c)} P(x)= 0 .
\end{align}
Using a change of variables and functions
\begin{align}
  \label{eq:5}
  \tilde{R}(x)&= e^{\frac{v}{2D}(x - x_c)}P(x)  \quad ,\\
  z&= \frac{2}{m} e^{-\frac{a}{2}(x - x_c)} \quad, \\
  R(z)&= \tilde{R}(x)\quad ,
\end{align}
we can rewrite Eq. \ref{eq:2} as 
\begin{align}
\label{eq:6}
z^2 \partial_z^2 R(z) + z \partial_z R(z) - \left[ \left( \frac{v}{a D} \right)^2 + z^2 \right]R(z)&= 0 \quad ,
\end{align}
where
\begin{align}
  \label{eq:37}
  a&\equiv \alpha \frac{x_c^{\alpha - 1}}{T}\\ \label{eq:mDef}
    m^2& \equiv a^2 D\dot{\gamma}\tau e^{\frac{x_c^{\alpha}}{T}}  \quad .
\end{align}

Eq. \ref{eq:6} is the modified Bessel equation and the solution, which vanishes for $x \to -\infty$, reads
\begin{align}
\label{eq:approxPx}
  P(x) &= C e^{-\frac{v}{2D}(x-x_c)} K_{\frac{v}{a D}}\left( \frac{2 e^{-\frac{a}{2}(x- x_c)}}{m}\right) \quad ,
\end{align}
where $K_\lambda(x)$ is the modified Bessel function of the second kind, of order $\lambda$. Finally, we can determine the integration constant $C$ and parameter $m$ by requiring that $P(x) \to P_0(x)$ as $T\to 0$. In the limit $T\to 0$ and for $x > x_c + \epsilon, \epsilon > 0$, the lowest order terms in the series representation of the Bessel function are:
\begin{align}
  \begin{split}
\label{eq:4}
  K_{\frac{v}{a D}}\left(\frac{2 e^{-\frac{a}{2}(x - x_c)}}{m}\right)&\approx \frac{1}{2}\Gamma \left( \frac{v}{a D} \right)m^{\frac{v}{a D}}e^{\frac{v}{2D}(x - x_c)} \\
  &+ \frac{1}{2}\Gamma \left( -\frac{v}{a D} \right)m^{-\frac{v}{a D}}e^{-\frac{v}{2D}(x - x_c)} \quad ,
  \end{split}
\end{align}
where $\Gamma$ is the gamma function. We assume that $\dot{\gamma}\exp(x_c^{\alpha}/T)$ does not depend exponentially on $T$ so that higher order terms are negligible when $T \to 0$. Equating $P(x)$ with $P_0(x)$ in the limit $T\to 0$ yields
\begin{align}
  \label{eq:34}
  C&=  \frac{2m^{-\frac{v}{a D}}}{v_c\Gamma \left( \frac{v}{a D} \right)}  \quad ,\\
  \label{eq:m}
m&= e^{\overline{\gamma}}\quad ,
\end{align}
where $\overline{\gamma}$ is the Euler-Mascheroni constant. With these expressions Eq. \ref{eq:approxPx} provides a solution of $P(x)$ in the vicinity of $x_c$.

Finally, from Eq. \ref{eq:mDef} and Eq. \ref{eq:m}, the thermal flow in the low temperature limit $T^{1/\alpha} \ll \Sigma_c - \Sigma$ is 
\begin{align}
\label{eq:50}
\dot{\gamma}= \frac{e^{2 \overline{\gamma}}}{\alpha^2 D\tau} T^2 \left(\frac{\Sigma_c - \Sigma}{\frac{1}{2} + D}\right)^{-2(\alpha - 1)}e^{-\frac{1}{(\frac{1}{2} + D)^{\alpha}}\frac{(\Sigma_c - \Sigma)^{\alpha}}{T}} \quad .
\end{align}
The thermal flow is exponentially small in $(\Sigma_c - \Sigma)^{\alpha}$, consistent with \cite{Johnson2005, Chattoraj2010}. 
It can be written in scaling form
\begin{align}
\label{eq:58}
\dot{\gamma}\sim T^{\frac{2}{\alpha}} f_{HL} \left( \frac{\Delta \Sigma}{T^{1/\alpha}}\right) \quad ,
\end{align}
where $\Delta\Sigma\equiv \Sigma - \Sigma_c$ and  $\lim _{y\to -\infty} f_{HL}(y) = |y|^{2(1 - \alpha)}\exp{(-c |y|^{\alpha})}$ with $c= 1/(1/2 + D)^{\alpha}$. Since the flow exponent $\beta= 2$ in HL model this scaling form is consistent with the prediction $\psi= \beta/\alpha$ in \cite{Middleton1992}.

\begin{figure}[ht!]
\centering
\includegraphics[width=.48\textwidth]{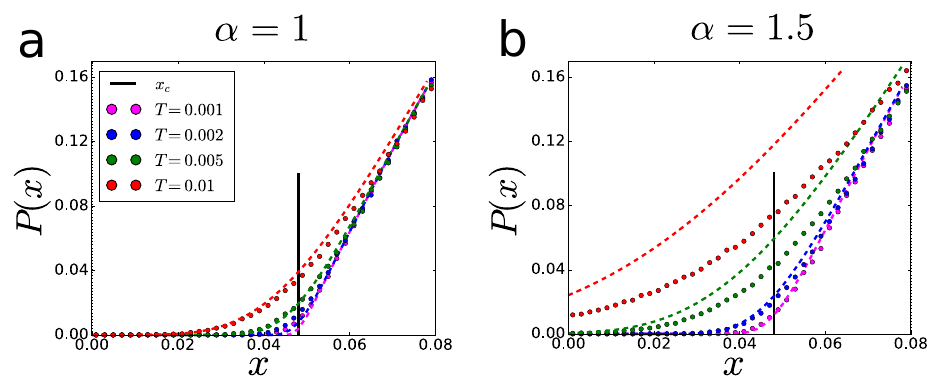}
\caption{Distribution $P(x)$ measured in HL model simulations (circles) compared to the analytical approximation Eq. \ref{eq:50} (dashed lines) in the vicinity of the gap $x_c$, indicated by a black line. a) For $\alpha = 1$  simulations and theory agree well at all measured temperatures. b) For $\alpha = 1.5$ the agreement is good at the lowest measured temperatures but it becomes significantly poorer at higher temperatures, as expected since the linearisation of thermal activation function is not a valid approximation.}
\label{fig:Px}
\end{figure}

\begin{figure}[ht!]
\centering
 \includegraphics[width=.41\textwidth]{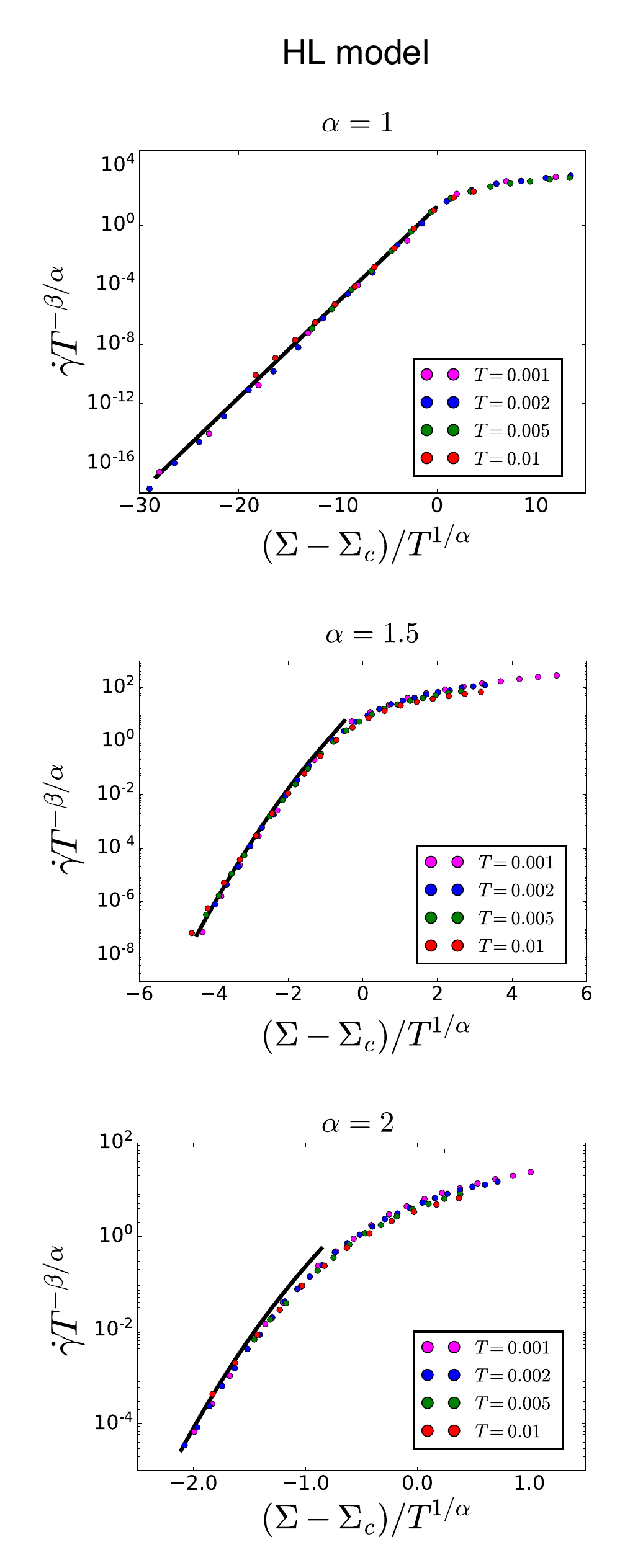}
\caption{Strain rate measured in numerical simulations using the HL model, for three different values of $\alpha = 1, 1.5, 2$. When scaled according to the proposed scaling law Eq. \ref{eq:scalingHypo}, the strain rates collapse. In addition, we find the analytical solution Eq. \ref{eq:50} (black solid line) is in excellent agreement with the simulations in the regime $(\Sigma - \Sigma_c)/T^{1/\alpha} \ll -1$ for which the solution was derived. }
\label{fig:scalingMF}
\end{figure}

\begin{figure}[ht!]
\centering
\includegraphics[width=.41\textwidth]{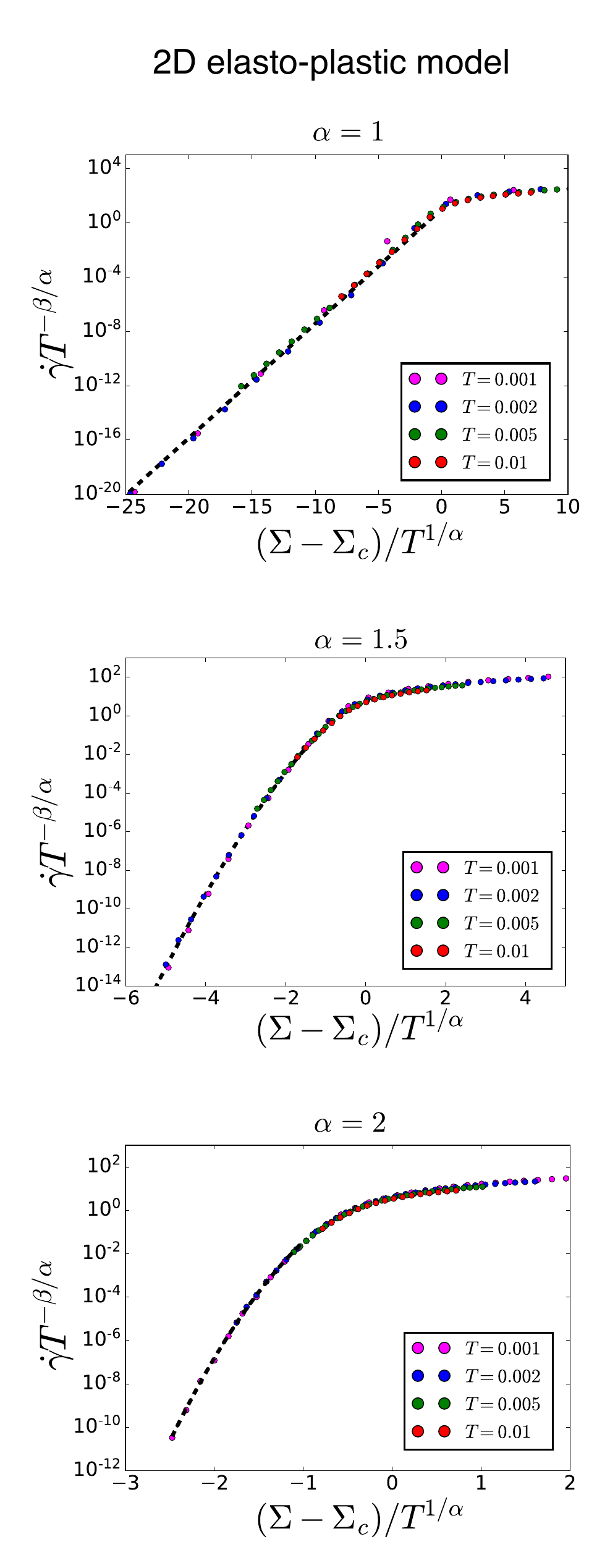}
\caption{Strain rate measured in numerical simulations using the two-dimensional elasto-plastic model, for three different values of $\alpha = 1, 1.5, 2$. The flow exponent for this model is $\beta= 1.51$ \cite{Lin2016a}. When scaled according to the proposed scaling law Eq. \ref{eq:scalingHypo}, the strain rates collapse.  The proposed scaling function in Eq. \ref{eq:scalingFunction}, shown in dashed black line, can account for the thermally activated regime.}
\label{fig:scaling2D}
\end{figure}

\begin{figure}[ht!]
\centering
\includegraphics[width=.48\textwidth]{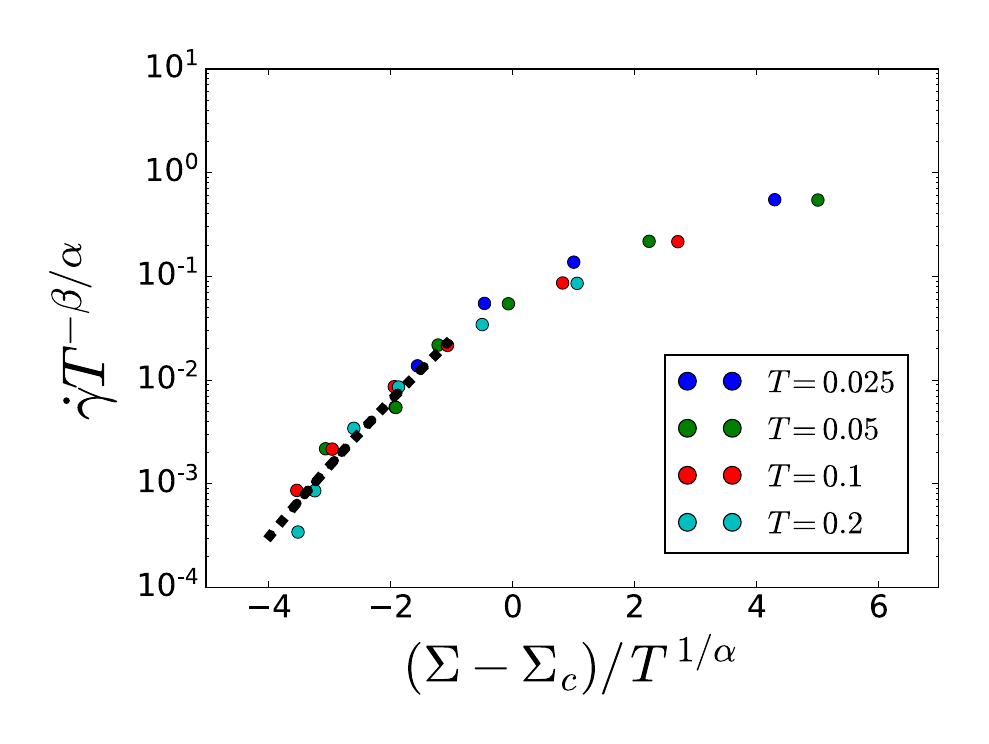}
\caption{Steady state strain rate measured in molecular dynamics simulations of two-dimensional glass extracted from \cite{Chattoraj2010}. We replotted the strain rate data as a function of $\Sigma - \Sigma_c$ on axes scaled according to the proposed scaling law Eq. \ref{eq:scalingHypo}. In this model particle interactions are smooth ($\alpha= 1.5$) and the flow exponent was measured to be $\beta= 2$. We find a good collapse of the data, indicating that Eq. \ref{eq:scalingHypo} holds beyond elasto-plasic models. To have comparable values of control parameter $(\Sigma - \Sigma_c)/T^{1/\alpha}$ with Figs. \ref{fig:scalingMF} and \ref{fig:scaling2D}, stress and temperature are normalised by shear modulus $\kappa$ and potential barrier scale $B$, respectively, using the reported values \cite{Chattoraj2010}. The proposed scaling function in Eq. \ref{eq:scalingFunction} is shown as dashed black line.}
\label{fig:MD}
\end{figure}

\section{Scaling law}
We propose the scaling form in Eq. \ref{eq:58} to hold in finite dimensional systems as
\begin{align}
\label{eq:scalingHypo}
\dot{\gamma}\sim T^{\psi} f \left( \frac{\Delta \Sigma}{T^{1/\alpha}}\right) \quad ,
\end{align}
where $\psi= \beta/\alpha$ and $\beta$ is the athermal flow exponent $\dot{\gamma}\sim \Delta \Sigma^{\beta}$. This form is the same as the one suggested by Fisher for thermal flow in depinning \cite{Fisher1985}. This scaling hypothesis assumes a characteristic stress scale ${|\Delta\Sigma| \sim T^{1/\alpha}}$, set by the activation $e^{-x^{\alpha}/ T}$ in the vicinity of the transition. The thermal rounding exponent $\psi= \beta/\alpha$ then follows by considering the athermal limit ${T^{1/\alpha} \ll \Delta \Sigma}$ in the vicinity of the transition. In this limit we conclude that $\lim_{y\to \infty}f(y) \sim y^{\beta}$, and therefore $\psi= \beta/\alpha$, in order to match the athermal flow.

Moreover, we propose the form of the scaling function $f$ in the thermally activated flow regime $\Sigma_c - \Sigma \gg T^{1/\alpha}$ we consider a system in the limit $T\to 0$ with a finite gap $x_c$. 
In this limit blocks become unstable in the vicinity of $x_c$ and the potential barrier in the activation function can be expanded to the first order around the gap $x_c$ as in Eq. \ref{eq:2}: $1/\tau \exp{(-x_c^{\alpha}/T)}\exp{(-\alpha x_c^{\alpha-1}(x - x_c)/T)}$. The first factor can be interpreted as a new time-scale $\tau(x_c)= \tau \exp{(x_c^{\alpha}/T)}$, and the second factor as a new effective activation function $\exp{(-(x- x_c)^{\alpha_{\text{eff.}}}/T_{\text{eff.}})}$, with effective values $\alpha_{\text{eff.}}= 1$ and $T_{\text{eff.}}= T/(\alpha x_c^{\alpha-1})$.

Since in the limit $T \to 0$ the effective absorbing boundary is at $x_c$, we expect that in the vicinity of $x_c$ the distribution $P_0(x)= \lim_{T\to0}P(x)$ corresponds to an athermal $P(x)$ of a system at a critical stress. 
Therefore, in this state the system 
state will respond to adding a small amount of temperature in the same way as a system at the critical stress. The flow can thus be described by the thermal rounding exponent $\dot{\gamma} \sim T_{\text{eff.}}^{\beta/\alpha_{\text{eff.}}}/ \tau(x_c)$ leading to: 
\begin{align}
\label{eq:flowHypotesis}
  \dot{\gamma} &\sim x_c^{-\beta(\alpha - 1)} T^{\beta} e^{-c \frac{x_c^{\alpha}}{T}} \\
  &\sim |\Delta\Sigma|^{-\beta(\alpha-1)}T^{\beta}e^{-c \frac{|\Delta\Sigma|^{\alpha}}{T}}\quad .
\end{align}
The last relation stems from $x_c \sim |\Delta\Sigma|$, i.e. the existence of a unique vanishing stress scale at $\Sigma_c$, and $c$ is a positive parameter. 
Therefore, the scaling function in the thermally activated regime reads:
\begin{align}
\label{eq:scalingFunction}
  f(y \ll -1) \sim |y|^{\beta(1 - \alpha)}e^{-c|y|^{\alpha}} \quad .
\end{align}

\section{Numerical tests}

To test the analytical results obtained in the HL model and the proposed finite dimensional scaling form for  $\dot{\gamma}$ we perform numerical simulations using HL and 2D elasto-plastic models, see the Appendix for details. 

\subsection{H{\' e}braud-Lequeux model}

We first compare the analytical approximation of the density $P(x)$ in the vicinity of $x_c$ in Eq. \ref{eq:approxPx} to the one obtained in HL model simulations. We find a good agreement between simulations and the analytical result for $\alpha= 1$ at all tested temperatures, see Fig. \ref{fig:Px}a. For $\alpha= 1.5$ the analytical approximation fails at higher temperatures, but a good agreement is recovered at lower temperatures, see Fig \ref{fig:Px}b. This is expected since the analytical solution was obtained assuming $T^{1/\alpha} \ll x_c$, which does not hold at higher temperatures, and consequently the linearisation of $x^{\alpha}$ is not justified. 

We next compare the analytical prediction in Eq. \ref{eq:50} and the proposed general scaling form Eq. \ref{eq:scalingHypo} with the results of HL and 2D elasto-plastic model simulations for $\alpha= 1, 1.5, 2$. We find that the strain rate measured at different temperatures collapses when the axes are scaled according to the proposed scaling form, see Fig. \ref{fig:scalingMF}. The analytical prediction of the mean-field strain rate given by Eq. \ref{eq:50} is shown as a black line in Fig. \ref{fig:scalingMF}, and it is in good agreement with numerical simulations at low temperatures in the thermally activated regime ${\Sigma_c - \Sigma \gg T^{1/\alpha}}$.

\subsection{2D elasto-plastic model}

We further test the generalised scaling form in Eq. \ref{eq:scalingHypo} using elasto-plastic simulations in two dimensions for ${\alpha= 1, 1.5, 2}$. The flow exponent measured in \cite{Lin2016a} for this model is $\beta= 1.51$. We find that the steady state flow rate collapses on a single curve when axes are scaled according to Eq. \ref{eq:scalingHypo}, see Fig. \ref{fig:scaling2D}. Furthermore, the scaling function Eq. \ref{eq:scalingFunction}, represented by a black dashed line, is consistent with simulation data for all three values of $\alpha$.
\subsection{Molecular dynamics}

Finally, to further verify the generality of the proposed scaling law, we extract strain rate curves obtained by molecular dynamics simulations of two-dimensional glass in ref. \cite{Chattoraj2010}. The good collapse of data when the axes are scaled according to Eq. \ref{eq:scalingHypo} is shown in Fig. \ref{fig:MD}. The scaling function we propose in Eq. \ref{eq:scalingFunction}, shown as dashed black line, is consistent with the data. Note that we normalized stress and temperature by shear elastic modulus $\kappa$ and potential barrier scale $B$ using the values provided in \cite{Chattoraj2010}.

\section{Discussion}
We have derived an approximation of thermally activated steady state strain rate in HL model of amorphous solids. We confirmed the validity of this expression with numerical simulations and generalised this result to a general scaling law for the steady state strain rate. We find that the proposed scaling form collapses both strain rate data from a two-dimensional elasto-plastic model and from molecular dynamics simulations \cite{Chattoraj2010}. Our results support that the thermally activated flow of amorphous solids can be described by a simple scaling law dependent only on the flow exponent $\beta$ and a parameter $\alpha$ reflecting properties of the disordered potential.

It is interesting to note that values of $\alpha$ different from $1.5$ have practical applications. For example, in cellular materials such as epithelial tissues or dry foams $\alpha = 2$ \cite{Popovic2020}. While thermal fluctuations are usually negligible in foams, mechanical noise from active processes in tissues can be a relevant factor in tissue flow \cite{Bi2016, Matoz-Fernandez2017} and future research of yield stress behavior in biological tissues will be able utilize and test results presented here.

The scaling form of the steady state strain rate was originally proposed in the context of depinning, it seems to describe well the rounding of the yielding transition in amorphous solids. The similarity between the two transitions is therefore useful to motivate further research of the yielding transition. An interesting research direction will be to study the low stress regime $\Sigma \ll \Sigma_c$. In the corresponding depinning regime of low forcing $f \ll f_c$, the interface velocity grows with $\ln{v} \sim -f^{-\mu}$. The exponent $\mu$ is associated with a diverging length scale on which the interface has to reorganize to cross the effective potential barrier \cite{Nattermann1987, Chauve2000, Ferrero2017}.

Finally, in this work, we have studied the steady state flow where all information about the initial state of the material has been erased. However, amorphous solids can exhibit a complex transient flow characterized by an initial slowing down followed either by eventual arrest or by sudden fluidisation. This phenomenon has been studied in the athermal HL and elasto-plastic models \cite{Liu18a, Liu18b}. However, it is important to understand the transient flow of thermal materials, where the arrest scenario is not available, and previously sharp transitions are smoothed on the stress scale $\Delta\Sigma \sim T^{1/\alpha}$.

\begin{acknowledgements}
  We thank  E. Agoritsas for useful discussions. M.W. thanks the Swiss National Science Foundation for support under Grant No. 200021-165509 and the Simons Foundation Grant ($\#$454953 Matthieu Wyart).
  T.G. acknowledges support the Swiss National Science Foundation (SNSF)
by the SNSF Ambizione Grant PZ00P2{\_}185843.
\end{acknowledgements}

\appendix*

\section{Hebraux-Lequeux and elasto-plastic model simulations}

\subsection{Implementation}
We implement the two-dimensional elasto-plastic model on a periodic lattice of linear size $L= 142$ \footnote{System size is $N= L^{2}$.}, following the implementation we used in \cite{Popovic2018}. The elastic dipole propagator $G(r, \phi)$ is a periodic version of an infinite system propagator $G_0(r, \phi) \sim \cos{4\phi}/r^2$ and it is normalised so that $G(\vec{r}= 0) = -1$. The sum of stresses along each row and column of elements is preserved. To keep the sum of stresses in all rows and columns the same during the initialization of the stress distribution $P(\sigma)$ we first apply the dipole propagator with a random prefactor drawn from a normal distribution $\mathcal{N}(0, 0.4^2)$ at each lattice block and then normalize the stress at each block by the sum of the absolute values of the propagator on the periodic lattice.
The initial yield stress distribution $P(\sigma_Y)$ is a normal distribution $\mathcal{N}(1, 0.1^2)$ and redrawn each time the block fails. These choices ensure that no stress overshoot and no shear banding occurs during the transient loading period.

After a failure local stress in the block is drawn from a normal distribution $\sigma_{i, after}=\mathcal{N}(0, 0.1^2)$ which defines the stress change in the block $\delta\sigma_i$.

The HL model simulations, in which blocks have no spatial information, contain $N= 20000$ blocks for the strain rate measurement and $N= 50000$ blocks for the $P(x)$ measurement. After each plastic event the stress is changed in all other blocks by an amount drawn independently from a normal distribution $\mathcal{N}(0, 2D/N)$, with $D= 0.18$.

To simulate thermal activation after each failure we draw the time until the next failure in the system from a Poisson distribution that takes into account all $x$ in the system. Then, we draw randomly the failing block by weighting each block with its failure rate $\exp{-x^{\alpha}/T}$. In this way duration of a simulation is proportional to the plastic strain, independent of the strain rate.
 
\subsection{Data analysis}
To measure the steady state strain rate $\dot{\gamma}$ and the distributions $P(x)$ we begin recording the state of the system only after it underwent a plastic strain of $5$. The steady state strain rate is then measured by sampling the strain rate after every $n_s$ plastic events up to the system plastic strain of $15$, and then calculating the median \footnote{Note that at low temperatures sampling the strain rate over too large intervals becomes very susceptible to finite size effects due to the exponential dependence of activation time on $x$}. In HL model simulations $n_s= 100$ in all cases except $\alpha= 1, T= 0.001$ which required $n_s= 10$. In 2D elasto-plastic model simulations $n_s= 1000$ in all cases except $\alpha=1, T= 0.001$ which required $n_s= 100$.  The steady state distribution $P(x)$ is measured in a system of size $N= 50000$ at the imposed stress $\Sigma= 0.29$ for the values of $\alpha$ and $T$ reported in Figure \ref{fig:Px}.

Values of $\Sigma_c$ in 2D elasto-plastic model and HL model were estimated by collapsing the strain rate data in Figs. \ref{fig:scalingMF} and \ref{fig:scaling2D}. Note that this was also required in the HL model since the relation $\Sigma_{c}= 1/2 - D$ holds only in the thermodynamic limit, while in finite systems value of $\Sigma_c$ is slightly modified by finite size effects \cite{Lin2014}.

\bibliographystyle{apsrev4-1}
\bibliography{bib}

\end{document}